\documentclass[conference]{IEEEtran}
\IEEEoverridecommandlockouts

\usepackage{cite}
\usepackage{amsmath,amssymb,amsfonts}
\usepackage{algorithmic}
\usepackage{graphicx}
\usepackage{textcomp}
\usepackage{xcolor}
\def\BibTeX{{\rm B\kern-.05em{\sc i\kern-.025em b}\kern-.08em
    T\kern-.1667em\lower.7ex\hbox{E}\kern-.125emX}}
\def\bC{{\mathbf{C}}}
\def\bH{{\mathbf{H}}}
\begin{document}

\title{Brain Age Gap Progression in Individuals at Risk for Neurodegeneration}

\author{\IEEEauthorblockN{Jason Scheffel\IEEEauthorrefmark{1},
Saurabh Sihag\IEEEauthorrefmark{2}, and The PREVENT-AD Research Group}
\IEEEauthorblockA{\IEEEauthorrefmark{1}\textit{Department of Mathematics \& Statistics},
\textit{University at Albany, SUNY}, Albany, United States \\
jscheffel@albany.edu}
\IEEEauthorblockA{\IEEEauthorrefmark{2}\textit{Department of Electrical \& Computer Engineering},
\textit{University at Albany, SUNY}, Albany, United States \\
ssihag@albany.edu}
}

\maketitle

\begin{abstract}
Neuroimaging is a widely used modality for characterizing neurodegeneration in various health conditions. Brain age gap has emerged as a promising biomarker for monitoring brain health from neuroimaging data using machine learning algorithms. Recent studies have leveraged coVariance neural networks (VNN)-driven pipelines based on the advancements in graph signal processing for relatively transparent construction of brain age gap from neuroimaging data. In this paper, we focus on a longitudinal study of VNN-based brain age gap in conjunction with Alzheimer's Progression Score (APS) in individuals that are at a higher risk of neurodegeneration due to family history of Alzheimer's disease. APS is a multimodal marker of disease progression. The longitudinal analysis of APS and VNN-derived brain age gap metrics revealed that a higher risk of disease progression at baseline predicted a larger rate of increase in brain age gap in follow-up visits. Thus, our findings revealed potential stratification in the brain age gap modulated by clinical/biological risks of AD in the at-risk population.
\end{abstract}

\begin{IEEEkeywords}
Graph neural networks, neurodegeneration, biomarkers
\end{IEEEkeywords}

\section{Introduction}
Aging manifests biologically as various progressive physiological and cognitive changes~\cite{lopez2013hallmarks}. Recent years have seen an increased focus in the study of brain aging using machine learning algorithms because of their relevance to digital health and precision medicine~\cite{baecker2021machine}. A common objective of brain age prediction strategies is to derive an estimate of brain age from neuroimaging data and compare it with chronological age (time since birth) via the brain age gap, i.e., the difference between brain age and chronological age. A large age gap in the brain has been shown to be indicative of accelerated aging in various neurological diseases, implying a greater burden of the disease and the risk of mortality~\cite{jove2014metabolomics}. Subsequently, we use the notation $\Delta$-Age to refer to brain age gap. Prior works have demonstrated that biological changes manifest within the brain at a significant amount of time (months to decades) before the clinical onset of a neurodegenerative condition~\cite{sperling2011toward,benussi2022conceptual}. Hence, digital markers of brain health, such as $\Delta$-Age, have the potential to make a significant impact in understanding and detecting neurodegenerative phenomenon in earlier stages of the disease. In this paper, we focus specifically on a cohort with family history of Alzheimer's disease (AD)~\cite{tremblay2021open}. Individuals with family history of AD have been reported to be several times more likely to develop AD relative to general population~\cite{cannon2019relative}.

The main objective of $\Delta$-Age prediction in neurodegeneration is to detect or characterize `accelerated aging'~\cite{cole2017predicting,franke2019ten}. To this end, a common template to infer $\Delta$-Age consists of: (i) training a regression model to predict chronological age for a healthy population; and then (ii) applying the regression model to predict age for a cohort of interest (often characterized by neurodegenerative condition). A higher $\Delta$-Age or its associations with clinical or biological measurements reflective of neurodegeneration hold the most relevance for it to be a biomarker in cohorts with health conditions~\cite{lee2022deep,elhaik2022principal,sihag2023explainable}. Furthermore, existing studies have leveraged longitudinal analysis of $\Delta$-Age to demonstrate its association with increasing disease severity in disease populations~\cite{franke2012longitudinal, cole2020longitudinal} and mortality risk in healthy individuals~\cite{cole2018brain}.

The existing literature in this domain has overwhelmingly focused on achieving near perfect performance in step (i) above with impressive reported accuracies in predicting chronological age across heterogeneous populations of healthy individuals~\cite{bashyam2020mri,peng2021accurate,niu2020improved}. However, the empirical evidence from several recent studies has hinted that the utility of $\Delta$-Age as a biomarker may not necessarily be determined by the prediction accuracy of chronological age on healthy population~\cite{bashyam2020mri,jirsaraie2023systematic,sihag2023explainable, schulz2025brain}. Such methodological obscurities in $\Delta$-Age add challenges to its interpretation in individuals at risk of neurodegeneration or in the presymptomatic stage as the magnitude of $\Delta$-Age alone may not be significantly high or yield information about the underlying pathology. In this context, we leverage the principles of the recently proposed \textit{explanation-driven} pipeline for $\Delta$-Age prediction~\cite{sihag2023explainable, sihag2025disentangling} based on coVariance neural network (VNN) models~\cite{sihag2022covariance} to study $\Delta$-Age in at-risk individuals.

\subsection{coVariance Neural Networks}
VNNs have recently been studied as graph neural networks based on principles of graph signal processing that leverage the sample covariance matrix as the graph shift operator~\cite{sihag2022covariance,cavallo2024fair,sihagJSTSP,sihag2023explainable}. The convolution mechanism within VNNs takes the form of \emph{linear-shift-and-sum} operators~\cite{ortega2018graph} on the covariance matrix. Specifically, given a sample covariance matrix ${\bC} \in \mathbb{R}^{m\times m}$, the convolution operation in a VNN is denoted by a \underline{coVariance filter}, given by $\bH(\bC) = \sum_{k=0}^K h_k \bC^k$, where scalar parameters $\{h_k\}_{k=0}^K$ are the learnable parameters and commonly referred to as filter taps.

\noindent
{\bf VNNs for neuroimaging data analysis.} VNNs have shown promise in deriving explainable and anatomically interpretable construction of biomarkers of neurodegeneration in Alzheimer's disease~\cite{sihag2023explainable, sihag2025disentangling,sihag2024towards,sihag2022predicting}. VNNs offer two key features in the context of deriving biomarkers from neuroimaging data~\cite{sihag2025disentangling}: (i) anatomically interpretable biomarkers~\cite{sihag2023explainable}; and (ii) cross-validation across neuroimaging datasets curated according to other brain atlases without any re-training~\cite{sihagJSTSP}.

\subsection{Contributions}
In this paper, we leverage the brain age gap prediction pipeline based on VNNs to study longitudinal evolution of $\Delta$-Age in individuals with family history of AD. The brain age gap prediction framework derived using VNN leverages the principles of graph signal processing and inherits the key advantages in terms of anatomical corroboration and reproducibility of findings~\cite{sihag2025disentangling}. The dataset of at-risk individuals was part of the PREVENT-AD study~\cite{tremblay2021open}. Alzheimer's Progression Score (APS) was available for a subset of participants. APS is a multimodal metric that has been previously validated as an estimate of disease progression in presymptomatic populations~\cite{leoutsakos2016alzheimer}. Our main finding is summarized as follows:
\begin{itemize}
    \item Baseline neurodegeneration risk, as measured by APS, predicted the trajectory of brain-age gap, with higher initial risk driving faster progression of brain age gap and vice-versa.
\end{itemize}
The above finding was supported by comprehensive statistical analyses, which further revealed distinct stratifications of at-risk individuals according to rate of progression of brain age gap and baseline APS scores.

Prior works on VNN-driven $\Delta$-Age prediction has focused primarily on neurodegenerative conditions~\cite{sihag2025disentangling, sihag2022predicting,sihag2023explainable,sihagJSTSP,sihagisbi}. This paper provides a distinct evaluation than prior works, as it demonstrates the utility of VNN-driven $\Delta$-Age in understanding neurodegeneration in at-risk individuals.

\section{Methods and Materials}
\subsection{PREVENT-AD Dataset}
Data used in preparation of this article were obtained from the PRe-symptomatic EValuation of Experimental or Novel Treatments for Alzheimer's Disease (PREVENT-AD) Registered repository available at https://registeredpreventad.loris.ca. The PREVENT-AD participants were cognitively healthy and identified to be at-risk to developing AD due to parental or multiple-sibling history of AD or related dementias~\cite{cannon2019relative,tremblay2021open}. For each individual, we used the T1-weighted (T1w) magnetic resonance imaging (MRI) images from their baseline scan to derive cortical thickness metrics curated according to Destrieux or Desikan-Killiany brain atlases using CAT-12 pipeline~\cite{gaser2022cat}. We leveraged the longitudinal structure of the PREVENT-AD dataset to characterize temporal relationships between VNN-derived $\Delta$-Age and APS in at-risk individuals. The longitudinal cohort comprised 703 MRI visits from 343 participants, with follow-up sessions at approximately 3, 12, 24, 36, and 48 months after baseline, yielding visit counts of BL00 ($n=343$), FU03 ($n=65$), FU12 ($n=97$), FU24 ($n=81$), FU36 ($n=70$), and FU48 ($n=47$). In total, 103 participants contributed at least two visits suitable for longitudinal modeling. Among the 161 participants with baseline APS scores, 66 individuals had APS measurements at two or more timepoints (253 total repeated APS measurements), enabling longitudinal modeling of APS trajectories. Per-visit cortical thickness features were processed identically to the cross-sectional analyses (CAT-12 Destrieux parcellation, neuroCombat harmonization, VNN pre-trained on $598$ healthy individuals in OASIS-3~\cite{lamontagne2019oasis}). Longitudinal models indexed time continuously as years elapsed since each participant's baseline session.

\subsection{$\Delta$-Age Prediction}
\noindent
{\bf Preprocessing and Training.} The VNN model was trained on the cortical thickness features derived from T1w MRI images collected from 3.0 Tesla MRI scanners for the healthy control population in the publicly available OASIS-3 dataset~\cite{lamontagne2019oasis}. The healthy population from OASIS-3 dataset consisted of $598$ individuals (age = $68.44\pm 7.64$ years). The cortical thickness features derived from structural MRIs using Freesurfer 5.3 were available as part of this dataset.

\emph{Harmonization} was performed using the neuroCombat package~\cite{fortin2018harmonization} on the cortical thickness features to mitigate variance due to different processing pipelines in OASIS-3 and PREVENT-AD datasets. To avoid data leakage between OASIS-3 and PREVENT-AD datasets, we used cortical thickness features derived with CAT-12~\cite{gaser2022cat} from data obtained from the CamCAN repository (https://cam-can.mrc-cbu.cam.ac.uk/dataset/), based on the CamCAN study~\cite{shafto2014cambridge,taylor2017cambridge}. OASIS-3 and CamCAN datasets were harmonized first, and the resulting harmonization estimates were then applied to PREVENT-AD.

\begin{figure*}[t]
    \centering
    \includegraphics[width=0.70\linewidth]{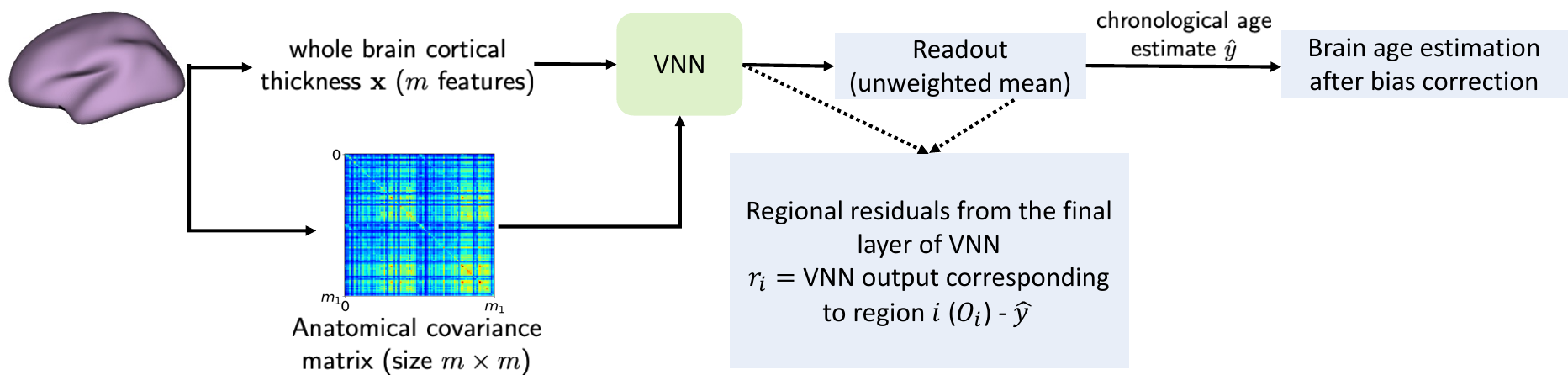}
    \caption{Brain age gap prediction pipeline using a VNN model~\cite{sihag2023explainable}. For the analysis in Section~\ref{refdyn}, the regional residuals $r_i$ are evaluated by subtracting the age estimate $\hat y$ from the VNN output corresponding to region $i$ (i.e., $O_i$).}
    \label{fig:vnn_pipeline}
\end{figure*}

VNN was trained to predict chronological age from 148 harmonized OASIS-3 cortical thickness features. The two-hidden-layer architecture used widths 63 and 39 with 14 and 22 filter taps, respectively, rectified linear unit (ReLU) activations, and was selected from 600 Optuna trials~\cite{akiba2019optuna}. OASIS-3 was split 90/10 into training and test sets. Training-set z-score parameters were applied to the test split, and the training covariance matrix was trace-normalized. Ten VNNs were trained with mean squared error (MSE) loss and Adam (learning rate 0.0008) using different random training/validation permutations, with each model using 90\% of the training split for training and 10\% for validation. The models achieved mean absolute error (MAE) $5.303\pm0.016$ years on training sets and $5.484\pm0.039$ years on the test set. PREVENT-AD $\Delta$-Age values were generated from one pre-trained VNN model following the setup in Fig.~\ref{fig:vnn_pipeline}, recently proposed based on advancements in graph signal processing~\cite{sihag2023explainable, sihag2025disentangling, sihagisbi}.

\section{Results}

The longitudinal analysis of the PREVENT-AD dataset helped characterize the temporal relationships between VNN-derived $\Delta$-Age and APS in at-risk individuals. Our preliminary results showed that APS score had an increasing trend over time (approximately 0.083 units per year on average across the considered population), while no such trend was observed for $\Delta$-Age gap. Cohort demographics, baseline replication, trajectory models, and growth-factor analyses are provided in Appendices~\ref{demo}--\ref{res3}.

\subsection{Baseline Values and Future Trajectories}

Despite the lack of average change in $\Delta$-Age over time, we examined whether baseline APS and baseline $\Delta$-Age were associated with individual trajectories by adding interactions between time and baseline values. For baseline APS predicting $\Delta$-Age trajectories, we used 71 participants with both baseline APS and longitudinal $\Delta$-Age data, comprising 353 $\Delta$-Age observations. Let $t$ denote years since baseline. We fit the linear mixed model:
\begin{align}
\Delta\text{-Age} &\sim t + \text{baseline\_APS} + t \times \text{baseline\_APS} \nonumber\\
& + \text{APOE4} + t \times \text{APOE4} + \text{baseline\_age} \nonumber\\
& + \text{sex} + \text{baseline\_education} + (1 + t \mid \text{participant})
\label{eq:baseline_aps_delta_age}
\end{align}
APOE4 was included in the model in \eqref{eq:baseline_aps_delta_age} as it has been shown to be a distinguishing factor for brain age in prior studies~\cite{lowe2016effect,fouquet2014imaging}. Baseline APS was associated with both the level ($\beta = 1.508$, $p < 0.001$) and the slope (time $\times$ baseline APS: $\beta = 0.126$, $p = 0.013$) of $\Delta$-Age. The level effect replicates the cross-sectional finding that higher APS at baseline is associated with higher $\Delta$-Age. Critically, the significant interaction term indicates that individuals with higher APS at baseline showed faster increases in brain age gap over time. Specifically, for each 1-unit increase in baseline APS, $\Delta$-Age increased by an additional 0.126 years per year of follow-up. This suggests that individuals who already have elevated multimodal markers of preclinical AD at baseline experience accelerated structural brain aging over time.

\subsection{Progression Heterogeneity}

To characterize heterogeneity in disease progression, we stratified the 66 participants with $\geq 2$ APS measurements into tertiles based on their APS slopes (extracted as empirical Bayes estimates (EBEs) from the APS trajectory model). The ``Fast'' progressors ($n=22$, highest tertile) had mean APS slope = 0.096 units per year, ``Medium'' progressors ($n=22$, middle tertile) had mean slope = 0.080 units per year, and ``Slow'' progressors ($n=22$, lowest tertile) had mean slope = 0.061 units per year. Baseline characteristics differed substantially across groups. Fast progressors had lower baseline APS (median = $-0.957$, range: $-1.496$ to 0.959) and lower baseline $\Delta$-Age (median = $-1.442$, range: $-3.539$ to 1.163), meaning their brains appeared younger than their chronological age on average. Slow progressors had higher baseline APS (median = 0.764, range: $-0.483$ to 2.117) and higher baseline $\Delta$-Age (median = 1.009, range: $-3.606$ to 5.672), meaning their brains appeared older than their chronological age. Medium progressors were intermediate on both measures.

\begin{figure}[!htbp]
\centering
\includegraphics[width=0.7\linewidth]{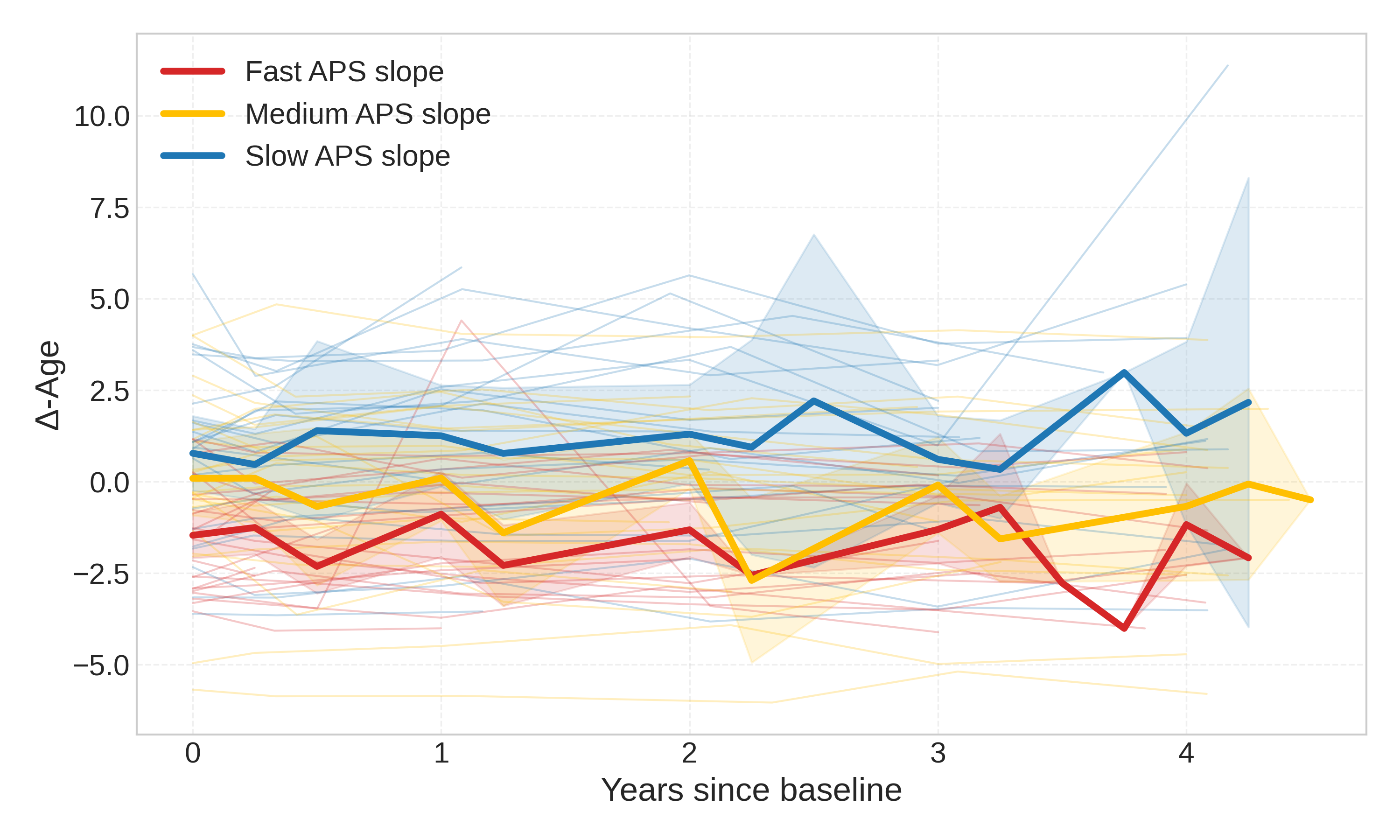}
\caption{Individual $\Delta$-Age trajectories stratified by APS progression rate. Participants were divided into tertiles based on their APS slopes (extracted as empirical Bayes estimates): Fast progressors (red, $n=22$, mean slope = 0.096 units/year), Medium progressors (orange, $n=22$, mean slope = 0.080 units/year), and Slow progressors (blue, $n=22$, mean slope = 0.061 units/year). Thin lines represent individual trajectories; thick lines show group means with shaded regions indicating 95\% confidence intervals. Fast APS progressors had lower baseline $\Delta$-Age (median = $-1.44$ years) compared to Slow progressors (median = 1.01 years), suggesting distinct disease trajectories within the preclinical AD spectrum.}
\label{fig:longit_heterogeneity}
\end{figure}

Examining $\Delta$-Age trajectories across groups (Fig.~\ref{fig:longit_heterogeneity}), at baseline (time = 0), Fast progressors had mean $\Delta$-Age = $-1.469$ years (standard error (SE) = 0.331, $n=22$), Medium progressors had mean = 0.091 years (SE = 0.516, $n=22$), and Slow progressors had mean = 0.775 years (SE = 0.520, $n=22$). At 2-year follow-up, these differences persisted: Fast progressors had mean $\Delta$-Age = $-1.314$ years (SE = 0.379, $n=17$), Medium progressors had mean = 0.575 years (SE = 0.391, $n=15$), and Slow progressors had mean = 1.299 years (SE = 0.689, $n=15$). This pattern suggests that progression profiles are fairly stable over this follow-up period, with individuals who start with rapid APS progression (but lower absolute levels) maintaining relatively lower $\Delta$-Age throughout, while those with slower APS progression (but higher absolute levels) maintain relatively higher $\Delta$-Age. This stability may reflect distinct disease trajectories or stages within the preclinical AD spectrum.

To validate the tertile classification, we performed k-means clustering on APS slopes (Table~\ref{tab:longit_clustering}). With $k=2$ clusters, the algorithm separated individuals into ``Low slope'' ($n=29$, mean APS slope = 0.0645) and ``High slope'' ($n=37$, mean slope = 0.0885) groups. The Low slope cluster contained all 22 Slow progressors plus 7 Medium progressors, while the High slope cluster contained all 22 Fast progressors plus 15 Medium progressors. With $k=3$ clusters, the algorithm produced groups with mean slopes 0.0601, 0.0801, and 0.0962, which aligned almost perfectly with the tertiles: the lowest cluster contained 20/22 Slow progressors, the middle cluster contained 22/22 Medium progressors, and the highest cluster contained 16/22 Fast progressors. This high concordance (overall agreement: 58/66 = 88\% for $k=3$) validates that the progression groups represent distinct data-driven clusters rather than arbitrary divisions, and that the tertile cutpoints reflect natural separations in APS slopes.

\begin{table}[!htbp]
\caption{Clustering validation: alignment between k-means clusters and tertile-based APS progression groups.}
\label{tab:longit_clustering}
\centering
\footnotesize
\textbf{(A) k-means with k=2 clusters}

\begin{tabular}{lccc|c}
\hline
\textbf{k-means Cluster} & \textbf{Slow} & \textbf{Medium} & \textbf{Fast} & \textbf{Total} \\
\hline
Low slope (0.065)  & 22 & 7  & 0  & 29 \\
High slope (0.089) & 0  & 15 & 22 & 37 \\
\hline
\textbf{Total}     & 22 & 22 & 22 & 66 \\
\hline
\multicolumn{5}{l}{\footnotesize Agreement: 44/66 (67\%)} \\
\end{tabular}

\textbf{(B) k-means with k=3 clusters}

\begin{tabular}{lccc|c}
\hline
\textbf{k-means Cluster} & \textbf{Slow} & \textbf{Medium} & \textbf{Fast} & \textbf{Total} \\
\hline
Low slope (0.060)    & 20 & 0  & 0  & 20 \\
Medium slope (0.080) & 2  & 22 & 6  & 30 \\
High slope (0.096)   & 0  & 0  & 16 & 16 \\
\hline
\textbf{Total}       & 22 & 22 & 22 & 66 \\
\hline
\multicolumn{5}{l}{\footnotesize Agreement: 58/66 (88\%)} \\
\end{tabular}

\scriptsize
\textit{Note:} Mean APS slopes (units/year) shown in parentheses. Columns represent tertile-based progression groups; rows represent k-means clusters. The k=3 solution shows superior concordance (88\% vs 67\%).
\end{table}

\subsection{Regional Dynamics}\label{refdyn}

We examined whether global $\Delta$-Age patterns were reflected regionally by fitting linear mixed models for each of the 148 cortical regions. For each region $i$, we modeled regional outputs ($O_i$, the VNN's final-layer representation) and regional residuals ($r_i$, the region-specific difference between the chronological age estimate and the VNN output; Fig.~\ref{fig:vnn_pipeline}). With $t$ denoting years since baseline, each regional model took the form:
\begin{align*}
\text{regional\_metric} &\sim t + \text{APOE4} + t \times \text{APOE4} \\
& + \text{baseline\_age} + \text{sex} + \text{baseline\_education} \\
& + (1 + t \mid \text{participant})
\end{align*}
Within each of these 148 regional models, both {time since baseline (years)} and the regional metric (whether $O_i$ or $r_i$) were z-score standardized prior to fitting to stabilize convergence. We applied both Bonferroni correction ($\alpha = 0.05/148 = 0.000338$) and false discovery rate (FDR) correction via the Benjamini-Hochberg procedure to control for multiple comparisons across the 148 regions. For regions showing significant time effects under either correction method, we conducted downstream analyses using ordinary least squares (OLS) regression: (1) testing baseline associations between regional metrics and APS (controlling for baseline age, sex, education), and (2) computing partial Pearson's correlations between regional slopes (extracted as EBEs) and global slopes ($\Delta$-Age slopes and APS slopes), controlling for baseline age, sex, education, and APOE4 status.

For regional outputs ($O_i$), 138 of 148 regions (93\%) showed statistically significant increases over time after Bonferroni correction, with 146 regions (99\%) significant under the more liberal FDR correction (Table~\ref{tab:longit_regional}). This widespread pattern indicates that the VNN detects structural changes across nearly all cortical regions over the follow-up period. These 138 Bonferroni-significant regions uniformly showed positive baseline associations with APS (138/138 regions, all $p < 0.05$ after Bonferroni correction for 138 tests), meaning that individuals with higher APS scores at baseline had higher regional output values in these regions---the VNN's learned representations of these brain regions already reflected greater pathology at baseline. All 138 regions also showed positive partial correlations with global $\Delta$-Age slopes (138/138 regions, all $p < 0.05$ after Bonferroni correction), meaning that individuals with faster global brain aging also showed faster increases in these regional outputs over time. This consistency between regional and global patterns validates that the VNN's region-specific representations track meaningfully with overall brain aging trajectories.

Regional residuals ($r_i$) exhibited greater heterogeneity in their temporal patterns. Of 148 regions, 71 showed statistically significant time effects after Bonferroni correction: 46 regions (31\%) showed increasing residuals over time while 25 regions (17\%) showed decreasing residuals. Under FDR correction, 124 regions showed significant time effects (71 increasing, 53 decreasing). This heterogeneity means that while global brain age increases (as reflected in the positive time effects for outputs), individual regions contribute differently---some regions show residuals that increase over time (meaning these regions increasingly contribute to higher brain age predictions as time progresses, suggesting regional vulnerability) while others show decreasing residuals (meaning these regions contribute less to brain age predictions over time, possibly reflecting relative preservation or compensatory mechanisms). Among the 46 regions with Bonferroni-significant increasing residual trajectories, 45 (98\%) showed positive baseline associations with APS ($p < 0.05$ after Bonferroni correction for 46 tests), meaning higher baseline APS was associated with higher residual values in these regions at baseline. All 46 (100\%) showed positive partial correlations with global $\Delta$-Age slopes ($p < 0.05$ after Bonferroni correction), confirming that individuals with faster global brain aging also showed faster increases in these regional residuals. Additionally, these regional residual slopes showed negative correlations with global APS slopes (14/46 regions significant at Bonferroni threshold, 45/46 at FDR threshold), reflecting the same inverse relationship seen at the global level: individuals with faster APS progression showed slower increases (or faster decreases) in regional residuals. This consistent pattern across multiple regions validates that the negative slope correlation observed globally is not a statistical artifact but rather a systematic feature that manifests anatomically across cortical regions.

\begin{table}[!htbp]
\caption{Summary of regional dynamics analyses across 148 cortical regions.}
\label{tab:longit_regional}
\centering
\setlength{\tabcolsep}{0.5pt}
\footnotesize
\begin{tabular}{lcccc}
  \hline
  \textbf{Metric} & \textbf{Significant} & \textbf{APS} &
  \textbf{$\Delta$-Age} & \textbf{APS} \\
  & \textbf{Time Effects} & \textbf{Association} & \textbf{Slope
  Corr.} & \textbf{Slope Corr.} \\
  \hline
  \multicolumn{5}{l}{\textit{Regional Outputs ($O_i$)}} \\
  Bonferroni & 138 incr. (93\%) & 138/138 (100\%) & 138/138 (100\%) & -- \\
  FDR & 146 incr. (99\%) & -- & -- & -- \\
  \hline
  \multicolumn{5}{l}{\textit{Regional Residuals ($r_i$)}} \\
  Bonferroni & 46 incr., 25 decr. & 45/46 (98\%) & 46/46 (100\%) &
  14/46 (30\%) \\
  FDR & 71 incr., 53 decr. & -- & -- & 45/46 (98\%) \\
  \hline
\end{tabular}
\end{table}

\section{Discussion}

This longitudinal analysis demonstrates that VNN-derived $\Delta$-Age serves as a dynamic biomarker in individuals at risk for Alzheimer's disease. Baseline APS was associated with both the level and rate of change in $\Delta$-Age (time $\times$ baseline APS: $\beta = 0.126$, $p = 0.013$). The negative correlation between $\Delta$-Age slopes and APS slopes ($r = -0.433$, $p = 0.0004$) likely reflects different disease stages: individuals with fast APS progression started with lower baseline APS and $\Delta$-Age, potentially representing earlier stages with rapid pathological accumulation, while those with slow APS progression had higher baseline values, potentially approaching plateau phases.

\appendices

\section{Demographics}\label{demo}
The BL00 cohort comprised 343 participants (age $63.62 \pm 5.06$ years, range 55.17--84.25; 243 females/100 males; education $15.43 \pm 3.35$ years, range 7--29; 131 $\text{APOE}\,\varepsilon4$ carriers). The longitudinal imaging subset ($\geq$2 MRI visits) included 103 participants (age $63.89 \pm 5.17$ years, range 55.25--84.25; 72 females/31 males; education $15.44 \pm 3.19$ years, range 7--24; 37 $\varepsilon4$ carriers).

Across visits, the dataset comprised 703 MRI sessions spanning BL00 (343), FU03 (65), FU12 (97), FU24 (81), FU36 (70), and FU48 (47). The longitudinal imaging subset contributed 463 scans (mean $4.50$ visits per person, range 2--6), extending to FU48 despite APS scores being limited to earlier follow-ups.

Baseline APS scores were available for 161 participants (age $63.66 \pm 5.50$ years, range 55.17--84.25; 116 females/45 males; education $15.12 \pm 3.37$ years, range 7--29; 62 $\text{APOE}\,\varepsilon4$ carriers). Of these, 66 provided longitudinal APS data (age $63.67 \pm 5.75$ years, range 55.25--84.25; 46 females/20 males; education $15.53 \pm 3.01$ years, range 9--23; 22 $\varepsilon4$ carriers), contributing 253 measurements across 3.83 visits per participant on average (range 2--4): BL00 (66), FU03 (65), FU12 (63), and FU24 (59). Across all APS visits, 348 observations from 161 participants ranged from $-1.61$ to $2.28$ (mean $0.02 \pm 0.87$).

APS scores were NAP-only: 161 of 162 NAP participants at BL00 had APS values, whereas PRE participants did not. Three individuals (1177880, 3230637, and 6794127) began in NAP but later transitioned to PRE, contributing baseline APS but no follow-up APS; therefore, APS analyses rely on NAP data while the broader MRI cohort spans both arms.

\section{Cross-Sectional Replication in Longitudinal Cohort}\label{res1}

\begin{figure}[!htbp]
\centering
\includegraphics[width=0.60\linewidth]{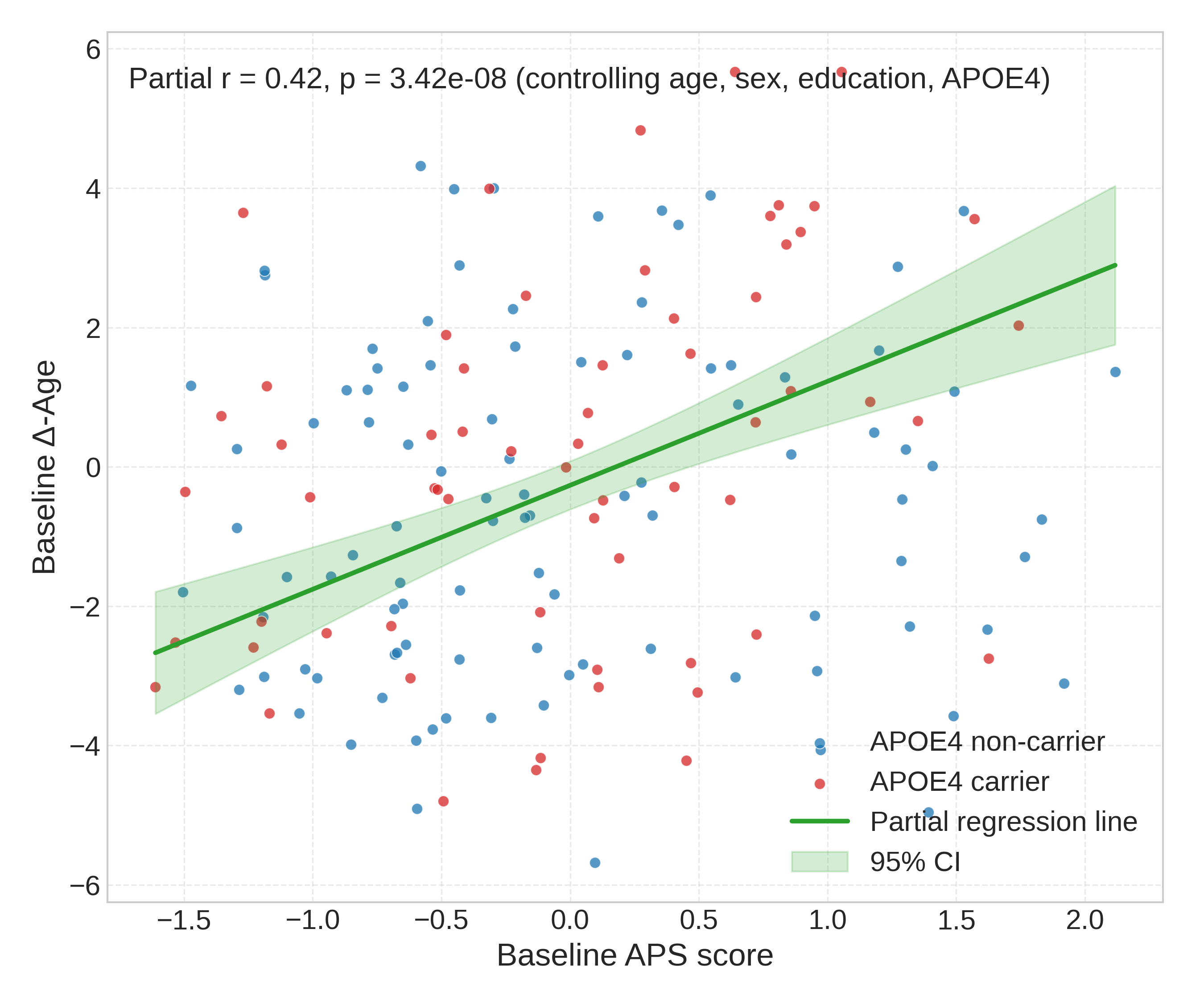}
\caption{Baseline association between $\Delta$-Age and APS among participants with APS measurements ($n=161$). The partial regression line and 95\% CI control for age, sex, education, and APOE4; points indicate APOE4 non-carriers (blue) and carriers (red). Higher APS was associated with higher $\Delta$-Age ($\beta = 1.492$, $p < 0.001$).}
\label{fig:longit_baseline}
\end{figure}

Using baseline data from all 343 participants, we replicated cross-sectional associations with OLS regression (Fig.~\ref{fig:longit_baseline}). In the demographic/genetic model for $\Delta$-Age (baseline age, sex, education, and APOE4), age was negatively associated with $\Delta$-Age ($\beta = -0.065$, $p = 0.016$), APOE4 showed a marginal positive association ($\beta = 0.533$, $p = 0.059$), and sex and education were not significant.

Among the 161 participants with baseline APS scores, an OLS model controlling for baseline age, sex, education, and APOE4 showed that higher APS was associated with higher $\Delta$-Age ($\beta = 1.492$, $p < 0.001$; adjusted $R^2 = 0.223$, 24.8\% variance explained). Age was negatively associated with $\Delta$-Age ($\beta = -0.239$, $p < 0.001$), males had lower $\Delta$-Age ($\beta = -1.197$, $p = 0.006$), and APOE4 was not significant ($p = 0.535$) once APS was included.

\section{Longitudinal Trajectories of APS and $\Delta$-Age}\label{res2}

\begin{figure}[!htbp]
\centering
\includegraphics[width=1\linewidth]{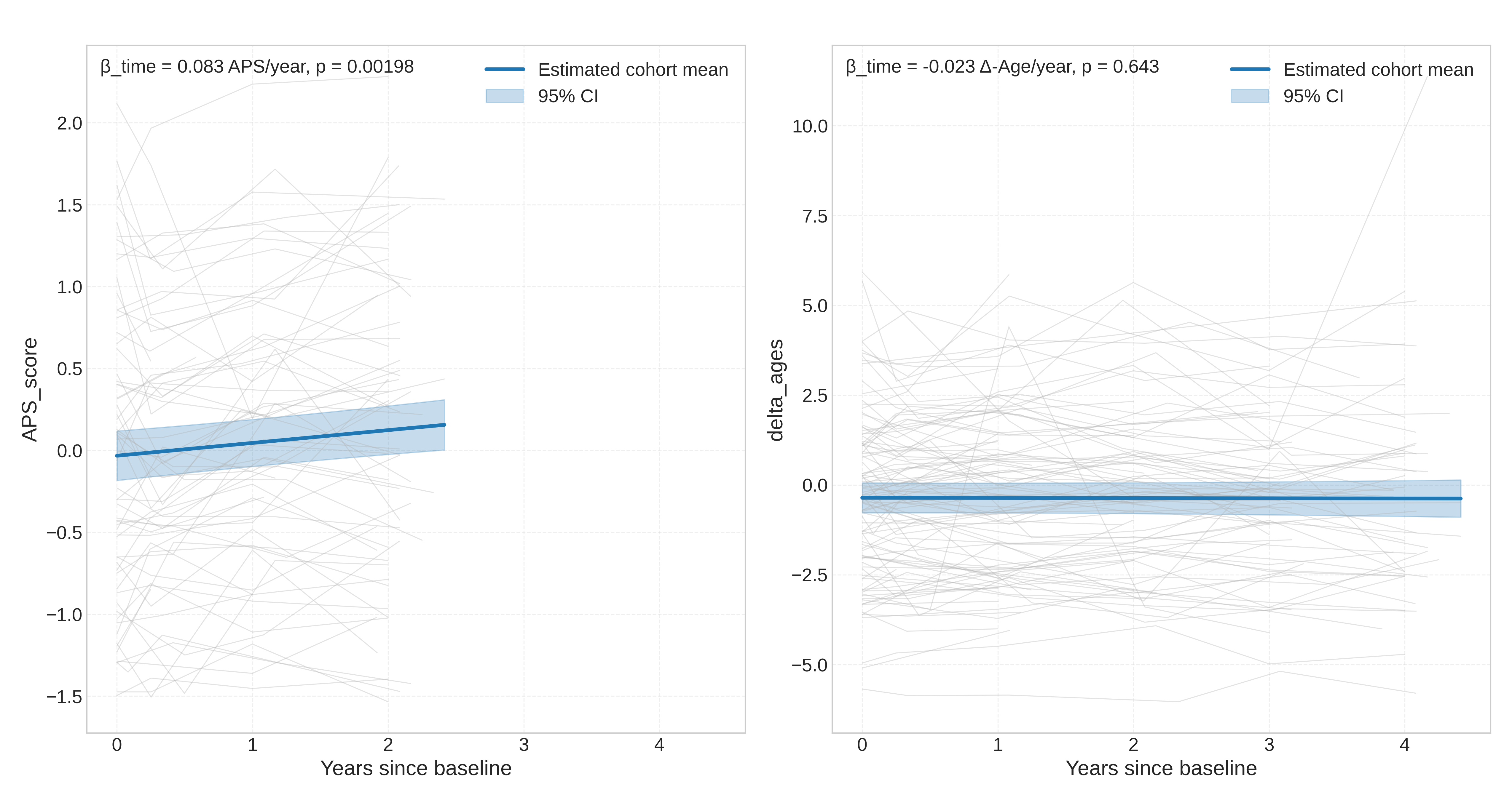}
\caption{Longitudinal APS and $\Delta$-Age trajectories. APS increased over time ($\beta = 0.083$ units/year, $p = 0.002$, $n=253$ observations from 66 participants), whereas $\Delta$-Age did not show population-level change ($\beta = -0.023$ years/year, $p = 0.643$, $n=463$ observations from 103 participants). Gray lines show individual trajectories; blue lines show cohort means with 95\% CIs.}
\label{fig:longit_trajectories}
\end{figure}

Temporal APS and $\Delta$-Age trajectories were modeled with random-intercept/random-slope linear mixed models (Fig.~\ref{fig:longit_trajectories}). For APS, 253 measurements from 66 participants with at least two APS timepoints were modeled as a function of years since baseline, APOE4, their interaction, baseline age, sex, education, and participant-level random intercepts/slopes. APS increased over time ($\beta = 0.083$ units/year, $p = 0.002$); APOE4 main effect ($p = 0.345$) and time $\times$ APOE4 interaction ($p = 0.767$) were not significant. Baseline age ($\beta = 0.073$, $p < 0.001$) and male sex ($\beta = 1.081$, $p < 0.001$) were associated with higher APS levels. In the analogous model for 463 $\Delta$-Age measurements from 103 participants with at least two MRI visits, $\Delta$-Age did not change significantly over time ($\beta = -0.023$ years/year, $p = 0.643$); APOE4 main effect ($p = 0.411$) and time $\times$ APOE4 interaction ($p = 0.530$) were also not significant, while baseline age was associated with $\Delta$-Age ($\beta = -0.083$, $p = 0.046$).

\section{Dynamics Between APS and $\Delta$-Age}\label{res3}

To characterize co-evolution of $\Delta$-Age and APS, individual slopes were extracted from the linear mixed trajectory models as empirical Bayes estimates (EBEs): $\Delta$-Age slopes from all 103 participants with $\geq 2$ MRI visits and APS slopes from the 66 participants with $\geq 2$ APS measurements. Partial Pearson correlations from OLS models controlled for baseline age, sex, education, and APOE4. Baseline APS correlated with $\Delta$-Age slopes ($r = 0.406$, $p = 0.0006$, $n = 71$), confirming that higher baseline APS predicted steeper $\Delta$-Age slopes. Among 66 participants with both slopes, $\Delta$-Age and APS slopes were negatively correlated ($r = -0.433$, $p = 0.0004$): faster APS progression corresponded to slower $\Delta$-Age progression, and vice versa. This may reflect different preclinical stages: fast APS progressors had lower baseline $\Delta$-Age (mean = $-1.46$ years) and APS (mean = $-0.96$), consistent with early rapid pathological accumulation before substantial structural brain changes, whereas slow APS progressors had higher baseline $\Delta$-Age (mean = $+0.78$ years) and APS (mean = $+0.76$), consistent with later preclinical plateauing.

\begin{figure}[!htbp]
\centering
\includegraphics[width=0.85\linewidth]{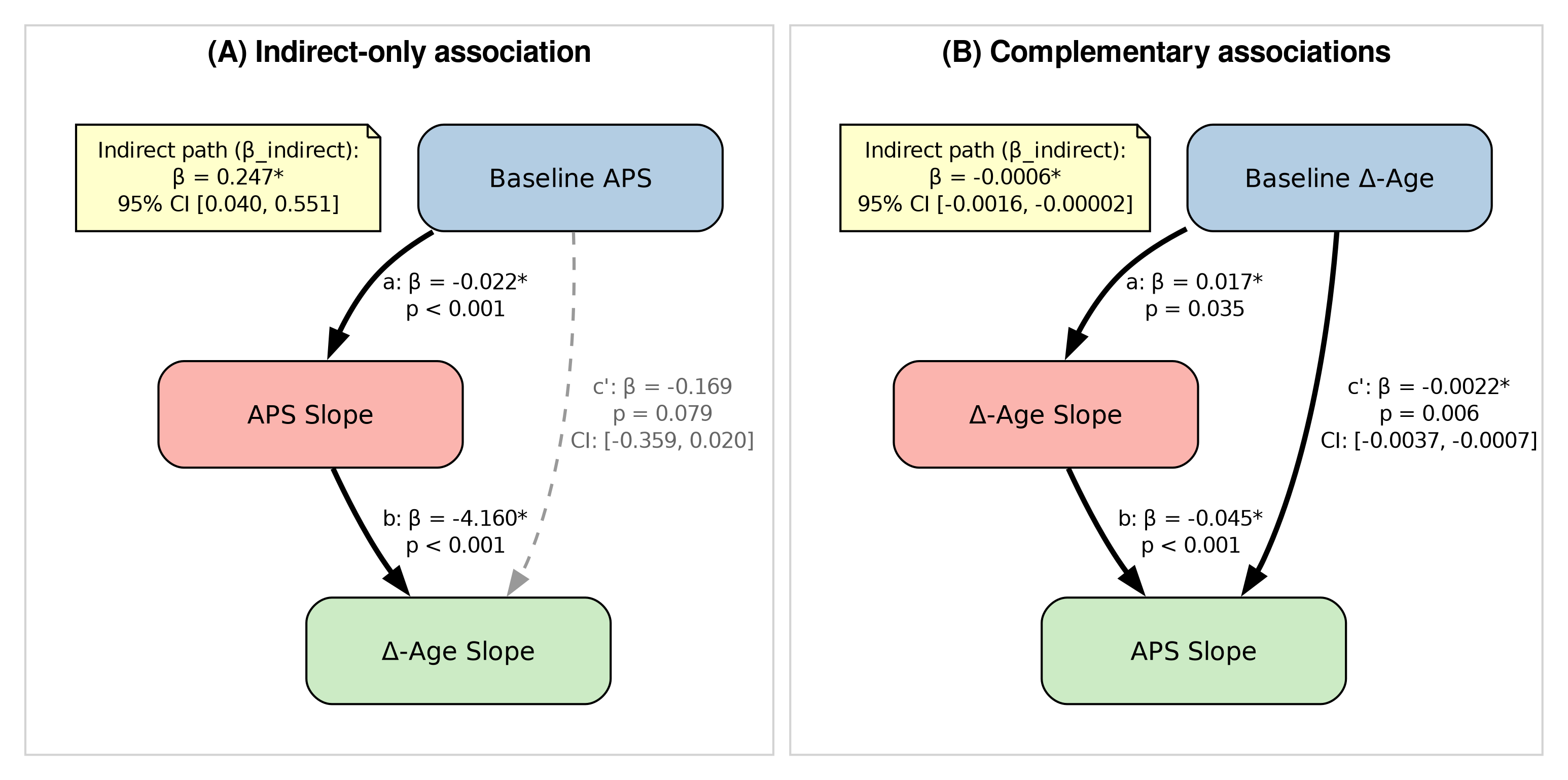}
\caption{Path analyses among baseline values and slopes. \textbf{(A)} Baseline APS showed a significant indirect path to $\Delta$-Age slopes through APS slopes ($\beta_{\text{indirect}} = 0.247$, 95\% CI [0.040, 0.551]) and a nonsignificant direct path ($\beta_{\text{direct}} = -0.169$, 95\% CI [$-0.359$, 0.020]). \textbf{(B)} Baseline $\Delta$-Age showed significant indirect ($\beta_{\text{indirect}} = -0.0006$, 95\% CI [$-0.0016$, $-0.00002$]) and direct ($\beta_{\text{direct}} = -0.0022$, 95\% CI [$-0.0037$, $-0.0007$]) paths to APS slopes. Models adjusted for baseline age, sex, education, and APOE4; concurrent slopes imply statistical, not causal, decomposition. Asterisks (*) indicate $p < 0.05$.}
\label{fig:longit_mediation}
\end{figure}

Path analyses among growth factors decomposed baseline-progression associations (Fig.~\ref{fig:longit_mediation}); because mediator and outcome slopes were estimated over the same interval, results are interpreted as statistical decompositions rather than causal mediation. Indirect effects used 10,000 bootstrap iterations, and all models adjusted for baseline age, sex, education, and APOE4. For baseline APS $\rightarrow$ APS slope $\rightarrow$ $\Delta$-Age slope, the indirect path was significant ($\beta_{\text{indirect}} = 0.247$, 95\% CI [0.040, 0.551], $p < 0.05$), whereas the residual direct path was not ($\beta_{\text{direct}} = -0.169$, 95\% CI [$-0.359$, 0.020], $p > 0.05$), indicating an \emph{indirect-only association}: higher baseline APS related to steeper $\Delta$-Age progression predominantly through APS progression.

For baseline $\Delta$-Age $\rightarrow$ $\Delta$-Age slope $\rightarrow$ APS slope, both the indirect path ($\beta_{\text{indirect}} = -0.0006$, 95\% CI [$-0.0016$, $-0.00002$], $p < 0.05$) and residual direct path ($\beta_{\text{direct}} = -0.0022$, 95\% CI [$-0.0037$, $-0.0007$], $p < 0.05$) were significant, indicating \emph{complementary associations} among growth factors.

Moderation analyses tested whether baseline APS and baseline $\Delta$-Age interacted to predict $\Delta$-Age and APS slopes. OLS models included mean-centered baseline APS, baseline $\Delta$-Age, their interaction, mean-centered baseline age and education, sex, and APOE4. The interaction was not significant for $\Delta$-Age slopes ($\beta = 0.0123$, $p = 0.185$) or APS slopes ($\beta = -0.0004$, $p = 0.159$), indicating additive rather than synergistic associations between baseline APS, baseline $\Delta$-Age, and future progression rates.

\section*{Ethics Statement}

This study used previously collected data from PREVENT-AD, OASIS-3, and CamCAN. No new human-subject recruitment, intervention, or data collection was performed by the authors. The original data collection procedures and consent procedures for these datasets were approved by their respective institutional ethics review boards or research ethics committees, and participants provided informed consent.

\section*{Acknowledgment}

Data collection and sharing for this project was provided by the Cambridge Centre for Ageing and Neuroscience (CamCAN). CamCAN funding was provided by the UK Biotechnology and Biological Sciences Research Council (grant number BB/H008217/1), together with support from the UK Medical Research Council and University of Cambridge, UK.

Data were provided in part by OASIS-3: Longitudinal Multimodal Neuroimaging: Principal Investigators: T. Benzinger, D. Marcus, J. Morris; NIH P30 AG066444, P50 AG00561, P30 NS09857781, P01 AG026276, P01 AG003991, R01 AG043434, UL1 TR000448, R01 EB009352. AV-45 doses were provided by Avid Radiopharmaceuticals, a wholly owned subsidiary of Eli Lilly.

\bibliographystyle{IEEEtran}
\bibliography{miccai_bib}

\end{document}